\pdfoutput=1
\RequirePackage{ifpdf}
\ifpdf % We are running pdfTeX in pdf mode
\documentclass[pdftex]{sigma}
\else
\documentclass{sigma}
\fi
\usepackage{mathrsfs}

\newcommand{\py}{{\hat{\pi}}}     % its conjugate momentum
\newcommand{\pyRI}{\hat{\pi}^{R,I}}

\newcommand{\fluc}[1]{(\Delta #1)^2}      % square of fluctuations
 
\newcommand{\cre}{\hat{a}^{\dagger}}    % creation

\newcommand{\ann}{\hat{a}}            % and annihilation operators

\newcommand{\tc}{\eta_{\vec{k}}^c}
\newcommand{\nk}{\vec{k}}
\newcommand{\dphi}{\delta \phi}

\newcommand{\bra}{\langle}
\newcommand{\ket}{\rangle}

\newcommand{\mpl}{M_{P}}

\begin{document}

\allowdisplaybreaks

\renewcommand{\thefootnote}{$\star$}

\renewcommand{\PaperNumber}{024}

\FirstPageHeading

\ShortArticleName{Novel Possibility of Nonstandard Statistics in the Inf\/lationary Spectrum}

\ArticleName{Novel Possibility of Nonstandard Statistics\\ in the  Inf\/lationary Spectrum of Primordial\\  Inhomogeneities\footnote{This
paper is a contribution to the Special Issue ``Loop Quantum Gravity and Cosmology''. The full collection is available at \href{http://www.emis.de/journals/SIGMA/LQGC.html}{http://www.emis.de/journals/SIGMA/LQGC.html}}}

\Author{Gabriel LE\'{O}N~$^\dag$ and Daniel SUDARSKY~$^\ddag$}

\AuthorNameForHeading{G.~Le\'{o}n and D.~Sudarsky}

\Address{$^\dag$~Department of Physics, University of Trieste, Strada Costiera 11, 34014 Trieste, Italy}
\EmailD{\href{mailto:gabriel.leon@nucleares.unam.mx}{gabriel.leon@nucleares.unam.mx}}

\Address{$^\ddag$~Instituto de Ciencias Nucleares, Universidad Nacional Aut\'onoma de M\'exico,\\
\hphantom{$^\ddag$}~M\'exico D.F. 04510, M\'exico}
\EmailD{\href{mailto:sudarsky@nucleares.unam.mx}{sudarsky@nucleares.unam.mx}}

\ArticleDates{Received August 31, 2011, in f\/inal form April 16, 2012; Published online April 21, 2012}

\Abstract{Inf\/lation is considered one of the  cornerstones of modern cosmology. However, the account of the origin of cosmic structure, as provided by the standard inf\/lationary paradigm, is not  fully satisfactory. The fundamental issue is the inability of the usual account to point out  the physical mechanism responsible for generating the inhomogeneity and anisotropy of our Universe, starting  from the exactly homogeneous and isotropic vacuum state associated with the early inf\/lationary regime.  We brief\/ly review this issue here together with the proposal to address this shortcoming in terms of a  dynamical collapse of the vacuum  state  of the inf\/laton f\/ield, which  has   been considered in previous works.
 The  main goal  of this  manuscript   is to discuss   certain  statistical aspects associated with the collapse and its implications
  in the primordial spectrum, particularly those connected with the possible  appearance of a novel type  of  unusual correlations.}


\Classification{83F05; 81T20; 81P05}

\renewcommand{\thefootnote}{\arabic{footnote}}
\setcounter{footnote}{0}

\section{Introduction}\label{Intro}

The inf\/lationary paradigm represents one of the  pillars of the $\Lambda$CDM cosmological model. It was  initially proposed  as a solution  to the classical naturalness problems of the the big bang models \cite{Guth_3,Guth_1,Guth_4,Guth_5,Guth_2, Guth}, but its impact became even more signif\/icant when it came to be regarded as the natural mechanism to account for the seeds of cosmic structure \cite{Mukhanov1990_3,Mukhanov1990_4,Mukhanov1990_1, Mukhanov1990,Mukhanov1990_5,Mukhanov1990_2}. However, as originally presented in \cite{Us}, the standard picture suf\/fers from a conceptual drawback: we know that our Universe is quantum mechanical and, thus, the classical descriptions must be regarded as nothing but short-hand and imprecise characterizations of complicated quantum mechanical states. The Universe that
we inhabit today is certainly very well described at the classical level by an inhomogeneous
and anisotropic classical state. That is, such description must be nothing but a concise and
imperfect description of an equally inhomogeneous and anisotropic quantum state. Consequently, if we want to seriously consider a theory, in which the early quantum state of the Universe was perfectly symmetric (the symmetry being the homogeneity and isotropy), then
we need to provide and explanation for  why the quantum state that describes our actual Universe lacks such symmetries. Since there is nothing in the dynamical evolution (as given by the standard inf\/lationary approach) of the quantum state that can break those symmetries,
then we are left with an incomplete theory. Therefore a natural inquiry arises: How does the actual inhomogeneities in the Universe we inhabit emerged?. The standard and phenomenologically successful accounts on this matter rely on the identif\/ication of the quantum f\/luctuations of certain observables associated to a homogeneous and isotropic Universe, with the averages over an ensemble of inhomogeneous Universes of their analogue classical quantities, i.e.\ the identif\/ication of quantum uncertainties and classical stochastic perturbations. Nevertheless, as discussed in \cite{Us,Short}, one can not avoid concluding that, using standard physics, there is no clear justif\/ication for such kind of identif\/ication. In fact, these shortcomings have been recognized by others in the literature \cite{Lyth,Mukhanov,Padmanabhan,Weinberg}.

The  idea  that  has been considered in previous  works \cite{Us,Sudarsky:2007tp,Sudarsky2011,Short, Sudarsky:2006zx}, as a possibility to deal with that problem, involves  adding  to the standard inf\/lationary
model the  hypothesis that  the   collapse of the  wave function is  an actual {\it physical} process that occurs  independently  of
external observers. The collapse hypothesis relies on the idea that something that ef\/fectively  might be described  in terms of such a collapse of the wave
function could have its origins in  the passage from the  atemporal regime of quantum gravity  to the classical   space-time
description underlining the  general theory of  relativity. That is, in going from one description to the other,  we  might be forced
to characterize   some aspects of the underlying physics in terms  of  sudden  jumps which are not compatible with the unitary Schr\"odinger evolution,
and  which    modify the  state of the Universe in an stochastic  way. Therefore, being capable of transforming  a condition that was
initially homogeneous and isotropic  into another one that is  not. It is worth mentioning that, in fact, conceptually similar ideas have been considered by many physicists concerned with the measurement problem in quantum mechanics \cite{Bassi, GRW, Pearle}, and the proposal that some quantum aspects associated to the gravitational interaction might lie at the root of this ef\/fect only serves to make the idea even more attractive \cite{Diosi,Diosi1,Penrose1, Penrose}.

The detailed discussion of the conceptual problems and those associated with the standard inf\/lationary paradigm  have  been  presented in  previous   works  by  some of  us and by others in \cite{Susana,Pearle, Us,Short}.  We  will not  reproduce  those arguments here.
On the other hand, a  detailed  discussion   of  the statistical aspects concerning both our modif\/ied  proposal involving the  quantum collapse of the wave function and as  well  as various aspects   of  the standard picture are  presented  in~\cite{Susana};  and  again we  will not  reproduce  most   of those arguments   here. In this paper our aim is to present one novel  possibility  regarding the statistical  aspects  which might  be present  in the  primordial spectrum.  It  would  correspond  to the  existence of non-trivial correlations between the random variables characterizing the collapse  of  certain modes and its statistical properties. In particular, these features would give raise  to a very particular  deviation
 from the  f\/lat shape  of the spectrum  of the primordial  f\/luctuations, and   thus  to  a  characteristic  signature  temperature anisotropies in the CMB. This is a crucial dif\/ference with the standard accounts, as  the distribution of the dif\/ferent  quantum f\/ield  modes has  previously been  assumed to be Gaussian, i.e.\ that all the  fundamental variables  are uncorrelated.

As we will show next, the  consideration of this  possibility  is enabled by  the fact  that the statistical aspects, in the collapse picture, are quite  transparent and clearly identif\/iable. In fact, the motivation for the work presented in this article, is the formal and rigorous analysis presented  in \cite{Alberto}. There, the collapse hypothesis is formally considered within the full framework of Quantum Field Theory in curved space-times, as this is the most reliable approach we can take in the absence of a complete theory of quantum gravity.

The paper is organized as follows. In Section~\ref{review}, we will review some of the  considerations contrasting  the collapse proposal and the standard approach.
We give  a brief  account  of way  we  think  that  these ideas  would f\/it into the  more standard  views of the quantum/gravity interface.
We also introduce the Semiclassical Self-consistent Conf\/iguration program which is the main motivation behind the results presented in this article. In Section~\ref{colapso},  we review the collapse models description for the inf\/lationary origin of the seeds of the cosmic structure. In Section~\ref{NG}, we present how the stochastic nature of the collapse proposal might induce new statistical features in the observable spectrum of anisotropies in the CMB. Finally, in Section~\ref{discusion}, we discuss our f\/indings. The conventions we will be using include a $(-,+,+,+)$ signature for the space-time metric. We will use units where $c=1$, but will keep the gravitational constant $G$ and $\hbar$ explicit throughout the paper.

\section{General framework of the collapse model}\label{review}

The collapse model for the  emergence of the seeds  of structure during  inf\/lation   was originally proposed in \cite{Us}, and since then has been further developed \cite{Adolfo,colapsosmuk,multiples,Gabriel}. In this section we will present a review of the ideas that we have explored and which are connected to the   signature  of novel statistical    features  explored in this paper.

Within the  standard approach \cite{Bartolo,Komatsu,Komatsu2001,Liguori, Yadav},  the possibility of  deviations  from simple statistics, is often referred  as  the study of   non-Gaussianity,
and is  a feature   thought to  be characterized  only in the   higher  order   statistical $n$-point functions. We  will discuss  a  novel type of deviation from the standard  statistics  that   is  suggested  by our  approach and that  could    lead to  modif\/ications  directly observable in  CMB    the spectrum.

In the  standard inf\/lationary paradigm, the perturbations of the f\/ield $\dphi$ and the perturbations of the curvature $\Psi$ are treated as standard quantum f\/ields\footnote{In fact they are both  treated as part of  a unif\/ied  f\/ield~$v$.} evolving in a classical quasi-De Sitter background space-time.
The statistical aspects  involve  a  series of  identif\/ication between  quantum $n$-point functions,  statistical $n$-point functions  for  ensembles of universes  and   averages over
our universe. Those  issues  are often left unspecif\/ied  and     the   clear characterization of these aspects  in the  theoretical analysis  is   often  missing.
On the other hand, within the collapse proposal, it becomes  quite evident where the stochastic nature of the  problem resides. That is, the dynamical collapse of the wave function is  taken  to be a physical mechanism governed essentially by  the stochastic rules inherent to the Quantum Theory. Although, we of course  do not know  what is the physics behind the collapse, we can parameterize it (in the next section we will detail this parametrization) and  hope  to  learn  something  about  its  basic  characteristics. The important point is that, if the initial state (say the Bunch--Davies (BD) vacuum state)  collapses (or is reduced) to a  dif\/ferent state, which lacks the symmetries of the original one, and  in a manner  in which  nontrivial  correlations   might  be  generated, then the   energy momentum tensor of  this resulting  state would  lead,  not only to  the emergence of  anisotropies and  inhomogeneities, but  might  result in a  particular  type of signature appearing in the primordial spectrum of f\/luctuations as  observable   features. We will explain  in the next sections  how   such  primordial non-f\/lat spectrum could arise, with the  departures from f\/latness representing  the  novel signatures we  have  been discussing.

\subsection{The semiclassical self-consistent conf\/iguration}\label{ssc}

As mentioned in the Introduction, the inf\/lationary scenario is a very peculiar setting where one can explore some of the fundamental aspects of modern physics. In particular, it provides us with a situation in which the ef\/fects of gravity and quantum mechanics are strongly coupled. However, we do not have a complete theory of Quantum Gravity (QG). In fact, one of the problems that arises when trying to unify General Relativity with Quantum Mechanics is the so called \emph{problem of time in quantum gravity}~\cite{isham}.

It is a known fact that the  proposals for  the canonical Quantization of Gravity leads to  timeless theories. This is simply because the lapse and shift vectors, naturally associated with the notion of time in General Relativity, disappear from the theory (the technical  reason is that the physical states of the theory need to satisfy the Hamiltonian and momentum constraints). The  Loop Quantum Gravity approach (LQG),  as the  modern  version  of such  ideas, does not escape from such  problem. Let us analyze in brief detail our view on how to deal  with it and its relation with our proposal (see \cite{Alberto,Sudarsky2011}).

In LQG,  one  starts  with the  characterization of gravity  in  terms  of the canonical variables given by he triad $E_i^a$ and the connection variables $A_a^i$. These variables are serve  to describe the geometry of a 3-spatial hypersurface, and characterize its embedding in a 4-dimensional (space-time) manifold.  At  the quantum level, however, one  passes  to  a description where the  basic  operators are  associated  with  holonomies and f\/luxes  $(\mathcal{U}_{\gamma},\mathcal{F}_{s})$   which  correspond  to  integrals of  the  connection and triads  over  closed  paths $\gamma$  and 2-surfaces  $s$ respectively.
The quantum   states  of the theory  are  thus  characterized  in terms   of these  variables, and the  recovery of  space-time  notions  from  this setting is  rather non trivial.
In fact  it seems  likely one  would need   special  kinds  of  semiclassical states  which  nevertheless should be  solutions of the constraints of the theory  including the Hamiltonian constraint. It is   still a challenge  to  construct these  states  and   the  full recovery  of space-time  from  them seems to  be   further away. However we can try to envision how this  process  would  look   when  such technical  dif\/f\/iculties  are overcome.

  Consider now, that we have incorporated into the theory,  along with the geometrical variables, some matter f\/ields, each  described
  by a  pairs of cannonical  variables $(\varphi_i, \pi_i)$.  In  most attempts to recover time, the crucial step  is the construction a quantum variable $\hat{T}(\varphi_i,\pi_i,\mathcal{U}_{\gamma},\mathcal{F}_{s})$, which might depend  on  both  the matter and gravitational degrees of freedom, and which  shouls play the role of as a  ``physical clock''. On the other hand, we  assume  we are given  a wave function for the conf\/iguration variables $\Phi(\varphi_1,\ldots,\varphi_n,{U}_{\gamma})$. Following the canonical quantization procedure, the wave function $\Phi$ must satisfy the Hamiltonian constrain, i.e.~$\hat{H}_\alpha \Phi(\varphi_1,\ldots,\varphi_n,{U}_{\gamma})=0$; $\alpha=0, 1, 2, 3,\ldots$. One can then obtain an ef\/fective wave function $\Psi$ for the remaining variables by using a projection operator $\hat{P}_{T,[t,t+\delta t]}$. The projector operator acts on the wave function $\Phi$ projecting it onto the subspace where the spectrum of
   $\hat{T}(\varphi_i, \pi_i,\mathcal{U}_{\gamma},\mathcal{F}_{s}) $, corresponding to the region between $t$ and $t +\delta t$, acquires some particular values.
    After  an ef\/fective wave  function $\Psi (t)$ has been obtained by  this procedure, one should be  able, in principle,  compute the expectations values of   geometrical  operators   like areas  and volumes  (as  well  as  others)   to  characterize  the spatial  metric and  extrinsic  curvature  associated  with the  ``hypersurface''  corresponding to the  selected value of $t$.  We  would do this  for  a large  enough  range of such values  and  obtain  data characterizing the   space-time in terms of  data for a slicing. In other words, the procedure described allows us to obtain an ``average'' 3+1 decomposition for the space-time, where the slicing would be identif\/ied with the hypersurfaces on which the geometrical quantities are given by the expectation values (calculated using the wave function $\Psi (t)$). Therefore, one would be able to use the standard lapse and shift functions to specify the space-time and the slicing. The  aspect   of this approach that we  want to focus  on,  is that, as  shown in \cite{Gambini1,Gambini}, the evolution of~$\Psi (t)$ would be given by a modif\/ied version of Schr\"odinger's equation including  departures from the standard unitary evolution.
 Thus it is not unnatural that an  ef\/fective   sort  of Schr\"odinger's equation, involving something  that looks, from the ef\/fective point of  view,  as  a collapse of the wave-function, could f\/ind its fundamental explanation in the setting we have proposed. In fact,  these types of modif\/ications have been previously proposed,  with dif\/ferent motivations, and  in  dif\/ferent  contexts,  by the Quantum Foundations community leading to the so called \emph{collapse models}
    \cite{Bassi,GRW, Pearle,Penrose1, Penrose}.

As  we do not have at  this  stage a complete  workable theory of QG, the framework under  which  our proposal has  been studied is directly
 that of semiclassical general relativity,
 which  we  have taken to  be just  an ef\/fective  description of limited  range of  applicability, and  which might  well  be  the only setting
 where standard  space-time notions  can  be  recovered  from an underlying  quantum gravity theory.
However, we  will assume that  its   applicability  includes the inf\/lationary cosmological   regime  which is the  situation of interest to us here.
 As  we  have  shown  in \cite{Alberto},  the nature of the  formal description one needs to  consider,  even
  though we do not take it to  be  in any sense  fundamental, is  not  as trivial as one might  initially  imagine.

In that approach,   one   considers that the Universe
can be  described,  by what we call a~\textit{Semiclassical Self-consistent Configuration} (SSC).
That is,  a space-time  geometry characterized by a~classical  space-time  metric  and  a standard quantum
 f\/ield theory  constructed  on that  f\/ixed space-time background,
together with a particular  state in that construction
such that the  semiclassical   Einstein  equations
hold.
In other words, we will say that
the set $\big\lbrace g_{\mu\nu}(x),\hat{\varphi}(x),\hat{\pi}(x)$, $\mathscr{H} \,\vert\,\xi\rangle\in\mathscr{H}\big\rbrace$
represents a
SSC if and only if $\hat{\varphi}(x)$, $\hat{\pi}(x)$ and $\mathscr{H}$  correspond to a quantum f\/ield theory
constructed over a space-time  with  metric~$g_{\mu\nu}(x)$ (as  described in, say~\cite{Wald1994}), and  the  state~$\vert\xi\rangle$ in~$\mathscr{H}$  is  such that
\begin{gather*}%\label{Mset-up}
G_{\mu\nu}[g(x)]=8\pi G\langle\xi\vert \hat{T}_{\mu\nu}[g(x),\hat{\varphi}(x),\hat{\pi}(x)]\vert\xi\rangle,
\end{gather*}
for all the points % $x=(t,\vec{x})$
in the space-time manifold.

Such  description  is  thought to  be  appropriate  in the   regime of interests  except  in those  times  when  a  collapse  occurs.
 The  resulting description is such that  when the initial    situation is homogeneous and isotropic, the    evolution can not   generate  any departure from that  symmetry. The  emergence of
inhomogeneities and  anisotropies requires therefore    something akin to   a collapse  process,  something   which in turn clearly  necessitates  a modif\/ied  description.
  In \cite{Alberto} we  deve\-lo\-ped  what we  consider  to be a natural prescription for the  description of   such   collapses,  and  we  analyzed in detail
    the  emergence through  such  process of  an  excitation of   a  single mode.

 The  proposal is as  follows:
 One considers, f\/irst within the Hilbert space associated to the  given SSC-i, that a transition $\vert\xi^{\textrm{(i)}}\rangle\to\vert\zeta^{\textrm{(i)}}\rangle_{\textrm{target}}$
``is about to happen'', with both $\vert\xi^{\textrm{(i)}}\rangle$ and $\vert\zeta^{\textrm{(i)}}\rangle_{\textrm{target}}$ in~$\mathscr{H}^{\textrm{(i)}}$.
Generically, the set $\{g^{\textrm{(i)}},\hat{\varphi}^{\textrm{(i)}},\hat{\pi}^{\textrm{(i)}}, \mathscr{H}^{\textrm{(i)}}\,\vert\,\zeta^{\textrm{(i)}}\rangle_{\textrm{target}}\}$ will
not represent a new SSC.
In order to have a sensible picture, our recipe is, as  discussed in \cite{Alberto},  to relate  this  state $\vert\zeta^{\textrm{(i)}}\rangle_{\textrm{target}}$ with another one
 $\vert\zeta^{\textrm{(ii)}}\rangle$ ``living'' in a new Hilbert space $\mathscr{H}^{\textrm{(ii)}}$ for which
$\{g^{\textrm{(ii)}},\hat{\varphi}^{\textrm{(ii)}},\hat{\pi}^{\textrm{(ii)}}, \mathscr{H}^{\textrm{(ii)}}\,\vert\,\zeta^{\textrm{(ii)}}\rangle\}$ is
an actual SSC. We denote the new SSC by SSC-ii.
Thus, f\/irst we need to determine the ``target'' (non-physical) state in $\mathscr{H}^{\textrm{(i)}}$ to which the initial state
is in a sense ``tempted'' to jump, and after that, one relates such target state with a corresponding  state
 in the Hilbert space of a new SSC, the SSC-ii.
Following our previous treatments on the subject
(see for instance~\cite{Us}), one then   considers that the
target state is chosen stochastically, guided by the quantum uncertainties of designated  f\/ield  operators,
evaluated on the initial state $\vert\xi^{\textrm{(i)}}\rangle$,  at the collapsing time.

 It is  natural to assume that  after the  collapse, the expectation values of the f\/ield and momentum operators  in each mode
  will  be related to the uncertainties  of the  pre-collapse  state (recall that the  expectation  values  in the vacuum state  are  zero).
  In the vacuum state,  the   operators   $\hat{y}_{\nk}$ and
$\hat{\pi}_{\nk}$ are individually characterized  by   Gaussian
wave functions centered at 0 with widths (uncertainties) $\fluc{\hat{y}_{\nk}}_0$ and
$\fluc{\hat{\pi}_{\nk}}_0$, respectively. The collapse  proposals  that we  have been considering  are based on the hypothesis that each mode  would  jump  to  a new state  where  the
  expectation value would be  determined  by both, the scale of the  uncertainties  and some random variable.  The idea
   is that the  collapse  would  be  a process taking place at a certain time,  and resembling  an  imprecise
   measurement,  which, of course, would be brought upon by  novel physics requiring no  observer or measuring apparatus.

A more precise   characterization of the  collapse process
  maintains the essence of such  prescription,  but now takes into account that the  change of the state  would need to be accompanied  by a
   change in the space-time metric  and that  would  require a  new  Hilbert space, and  a  new    quantum state in that Hilbert space.
  As  shown in \cite{Alberto},  the scheme  serves  to  characterize a complete  post-collapse  SSC (i.e.\  the  SSC-ii).
In fact,   regarding the identif\/ication between the two dif\/ferent SSC's involved in the collapse,
 we  took   in \cite{Alberto}  what turned  out to be  a  very  promising
prescription: Consider that the collapse takes place along a Cauchy hypersurface $\Sigma$.
A transition from the physical state $\vert\xi^{\textrm{(i)}}\rangle$ in $\mathscr{H}^{\textrm{(i)}}$ to the physical state
$\vert\zeta^{\textrm{(ii)}}\rangle$ in $\mathscr{H}^{\textrm{(ii)}}$ (associated to a certain  target \textit{non-physical} state
$\vert\zeta^{\textrm{(i)}}\rangle_{\textrm{target}}$ in $\mathscr{H}^{\textrm{(i)}}$) will occur  in a way that
 \begin{gather*}%\label{recipe.collapses}
_\textrm{target}\langle\zeta^{\textrm{(i)}}\vert \hat{T}^{\textrm{(i)}}_{\mu\nu}[g^{\textrm{(i)}},
\hat{\varphi}^{\textrm{(i)}},\hat{\pi}^{\textrm{(i)}}]\vert\zeta^{\textrm{(i)}}\rangle_{\textrm{target}} \big|_{\Sigma}=
\langle\zeta^{\textrm{(ii)}}\vert \hat{T}^{\textrm{(ii)}}_{\mu\nu}[g^{\textrm{(ii)}}, \hat{\varphi}^{\textrm{(ii)}},\hat{\pi}^{\textrm{(ii)}}]\vert\zeta^{\textrm{(ii)}}\rangle \big|_{\Sigma} ,
\end{gather*}
i.e.\  in such a  way  that   the expectation value of the energy momentum tensor, associated to the states $\vert\zeta^{\textrm{(i)}}\rangle_{\textrm{target}}$ and
$\vert\zeta^{\textrm{(ii)}}\rangle$ evaluated on the Cauchy hypersurface $\Sigma$, coincides. Note that the
 left hand side  in the expression  above
is meant  to be constructed from the elements of the SSC-i (although $\vert\zeta^{\textrm{(i)}}\rangle_{\textrm{target}}$ is not really {\it the state} of the SSC-i), while the right hand side correspond to quantities  evaluated  using the SSC-ii.

In the  situation of interest studied in \cite {Alberto},  the SSC-i  corresponded to the homogenous and  isotropic  space-time    characterized by $\Psi=0$ with the state of the quantum f\/ield corresponding to the BD vacuum. Meanwhile, the  SSC-ii  corresponded  to an excitation of  a single mode  (with  co-moving  wave number  $\vec{k}_0$), characterized  by a  Newtonian potential given  by $\Psi= F(\eta) \cos (\vec{k}_0 \cdot \vec x )$,
% (where $\Phi$  is a  phase),
and  a specif\/ic quantum state for the  inf\/laton  f\/ield. The energy momentum tensor for that  state is compatible with this  space-time  metric according to  the SSC recipe.

The point is that,  as  shown in  detail  in \cite{Alberto},  when the  SSC-ii  corresponds to  a potential  inclu\-ding  spatial  dependences   with  wavenumber $\vec{k}_0$
the normal  modes  of the   corresponding Hilbert space    $\mathscr{H}^{\textrm{(ii)}}$,  which  would  otherwise  be characterized  by the typical   spatial dependence~$e^{i \vec{k}\cdot\vec x}$,  and  would  be called  the $\vec{k}$  modes,   would  now  contain also  corrective  contributions   of the form~$e^{i( \vec{k}\pm \vec{k}_0)\cdot\vec x}$.

In turn, this implies   that  if we  want to  ensure that the  energy momentum tensor does  not lead to the excitation  of  other  modes  besides the $ \pm \vec{k}_0$ in the Newtonian potential, then the  state of the quantum f\/ield  needs to  be excited,  not only in the modes $\pm \vec{k}_0$,  but also  in the  modes~$ \pm  n \vec{k}_0$,  with~$n$ integer (with the degree  of excitation  decreasing with increasing~$n$). In those circumstances,  one can not longer  argue that  at the  time  when   these  higher  modes collapse, there  should be  complete  symmetry in the {\it a~priori}  statistical  distribution of the post-collapse   parameters  of the state.

 That is,   there seems to be ample  justif\/ication to assume that  the  value of the  random   parameter $x_{n \vec{k}_0}$ ($n\not = 1$)  would not  be  completely independent from the   value taken   by the random parameters  $x_{\vec{k}_0,y} $ and $x_{\vec{k}_0,\pi} $. Focussing for simplicity on the  modes,   which in view of the  prece\-ding   discussion one  expects to  be  more closely correlated, we  limit  hereafter consideration to the  case  $n=2$, and  characterize  the level of correlation by  an unknown  quantity  we label $ \varepsilon$.
  We  will show in detail   how these ideas  are implemented  in  Section~\ref{NG}.

We should note  here  we  will   not be using  at this point the  full  f\/ledged  formal  treatment developed  in~\cite{Alberto}.
  This is because, as  can be  seen there,  the  problem  becomes extremely cumbersome even in  the treatment of a single  mode. Thus, even though it  is in principle  possible to use  such  detailed  formalism   to treat   the  complete  set of relevant modes, when studying the  CMB spectrum  the  task  quickly becomes a practical impossibility.  We  will instead rely on the  less formal  treatments  we had  employed in previous works, and  retain from  the  results  of~\cite{Alberto}  only the   unusual  correlations  we  discussed above.
  Before doing so,  we  will brief\/ly review the basic  details, of  the   simple  version of our formalism. The reader should be warned that this is not (and can not expected to  be) a fully  self contained  work, as  doing that   would involve repeating extensive discussions  and  analysis   already  published  in the literature.

\section{The collapse model for the inf\/lationary f\/ield}\label{colapso}

The starting point of the   specif\/ic  analysis  is the same  as the standard picture, i.e.\ the action of a scalar f\/ield coupled to
gravity:
\begin{gather*}
%\label{eq_action}
S=\int d^4x \sqrt{-g} \left[ {\frac{1} {16\pi G}} R[g] - 1/2\nabla_a\phi
\nabla_b\phi g^{ab} - V(\phi)\right],
\end{gather*}
 where $\phi$ stands for the inf\/laton and $V$ for the
inf\/laton's potential.
 One then splits both, metric and
scalar f\/ield into a spatially homogeneous (background) part and an
in-homogeneous part (f\/luctuation), i.e.\ $g=g_0+\delta g$,
$\phi=\phi_0+\delta\phi$.

\looseness=-1
Note that  in the  full  f\/ledged  formalism  introduced  in \cite{Alberto},  the quantum description  is applied, not only to the inf\/laton
 mode  perturbation,  but  also to  the ``background  f\/ield''  which is  cha\-racte\-rized in terms of the zero mode  of a quantum f\/ield  theory construction on  the inf\/lating space-time.
  At the same time,  the  metric sector is  treated   exclusively at the  classical level as  corresponding to  the semiclassical  description we  have  relied on (for more on the  justif\/ication  for this see \cite{Alberto}). Here however, we  will return to the   simpler but more convenient   formalism    of previous treatments,  retaining as  we  have said only  the interesting  novel correlations  we  discussed. Thus,  here the zero mode of the  inf\/laton f\/ield  will be  treated  as a  classical   background f\/ield.

The background is  taken to  be the spatially f\/lat Friedmann--Robertson Universe  with line element
$ ds^2=a(\eta)^2\left[- d\eta^2 + \delta_{ij} dx^idx^j\right]$, and the homogeneous scalar f\/ield $\phi_0(\eta)$. The  evolution equations  for this  background are
scalar f\/ield equations,
\begin{gather*}
\ddot\phi_0 + 2 \frac{\dot a}{ a}\dot\phi_0 +
a^2\partial_{\phi}V[\phi] =0,
 \qquad
3\frac{\dot a^2}{a^2}=4\pi G  \big(\dot \phi^2_0+ 2 a^2 V[\phi_0]\big).
\end{gather*}
The  scale factor  solution corresponding  to the inf\/lationary era  of  standard inf\/lationary cosmo\-lo\-gy, written using a conformal time, is:
$a(\eta)=-1/[H_I^2(1-\epsilon)\eta]$
 with $ H_I ^2\simeq  (8\pi/3) G V$, $\epsilon$ is the slow-roll parameter which during the inf\/lationary stage is considered to be very small $\epsilon \ll 1$, thus $H_I \simeq {\rm const}$, and with the scalar  f\/ield $\phi_0$ in the slow roll regime, i.e.\ $\dot\phi_0= - ( a^3/3 \dot a)V'$.
  According to   the standard  inf\/lationary scenario, this era is followed by  a reheating period in which the Universe is repopulated with ordinary matter f\/ields,
   a regime that  then evolves towards a standard hot
 big bang cosmology regime leading up to the present cosmological time. The functional  form  of $a(\eta)$ during these latter periods changes, but we
  will ignore those details because most of the change in the value of $a$ occurs during the inf\/lationary regime.
  We  will set  $ a=1$ at the ``present  cosmological time'',  and assume that  inf\/lationary regime ends at a value of $\eta=\eta_R$, negative and very small  in absolute terms ($\eta_R \simeq -10^{-22}$ Mpc).

   Next,  we consider  the perturbations. We will ignore vectorial perturbations, and focus on the scalar and tensorial ones. First, we will focus on the scalar perturbations.  Working in  the  so called  longitudinal
    gauge, the perturbed metric is  written as:
\begin{gather*}
ds^2=a(\eta)^2\left[-(1+ 2 \Psi) d\eta^2 + (1- 2
\Psi)\delta_{ij} dx^idx^j\right],
\end{gather*}
 where $\Psi$  stands for the scalar perturbation usually known as
the \emph{Newtonian potential}.

The perturbation of the scalar f\/ield is  related to a perturbation of the energy momentum tensor, and
ref\/lected  into  Einstein's equations  which at lowest order  lead to the following   constraint equation for the  Newtonian potential:
\begin{gather}
\nabla^2 \Psi  =4\pi G \dot \phi_0  \delta\dot\phi= s   \delta\dot\phi,
\label{main3}
\end{gather}
where we introduced the abbreviation $s \equiv 4\pi G \dot \phi_0$.

Regarding  the f\/ield $\delta\phi$,
it is convenient  to  work with the  rescaled f\/ield variable  $y=a \delta \phi$ and its conjugate momentum $ \pi = \dot{y} - \frac{ \dot{a}}{a} y=  a \delta\dot\phi $.
We consider the quantum theory of this  f\/ield setting  the problem, for simplicity,  in  a f\/inite   box of side $L$,  which can be taken to $ \infty$  at the end of all calculations.  Thus  we  write  the   expansions  for
f\/ield  and  conjugate momentum opera\-tors~as
\begin{gather*}
\hat{y}(\eta,\vec{x})=
 \frac{1}{L^{3}}\sum_{\vec k}\ e^{i\vec{k}\cdot\vec{x}} \hat y_{\nk}
(\eta), \qquad \py(\eta,\vec{x}) =
\frac{1}{L^{3}}\sum_{\vec k}\ e^{i\vec{k}\cdot\vec{x}} \hat \pi_{\nk}
(\eta),
\end{gather*}
where the sum is over the wave vectors $\vec k$ satisfying $k_i L=
2\pi n_i$ for $i=1,2,3$ with $n_i$ integer and where $\hat y_{\nk} (\eta) \equiv y_k(\eta) \ann_{\nk} + y_k^*(\eta)
\cre_{-\nk}$ and  $\hat \pi_{\nk} (\eta) \equiv g_k(\eta) \ann_{\nk} + g_{k}^*(\eta)
\cre_{-\nk}$
with the usual  choice of modes:
 \begin{gather*}
 y_k(\eta)=\frac{1}{\sqrt{2k}}\left(1 - \frac{i}{\eta
k}\right)\exp(- i k\eta), \qquad
g_k(\eta)=-
i\sqrt{\frac{k}{2}}\exp(- i k\eta),
%. \label{Sol-g}
\end{gather*}
 which  leads to  what is known  as the Bunch--Davies  vacuum.

 Note  that according to the point  of  view  we  discussed in Section~\ref{ssc},   and  having at  this point the quantum theory for the relevant matter f\/ields, the ef\/fects  of the quantum f\/ields   on the geometrical variables  are codif\/ied  in the  semiclassical Einstein's equations.  Thus  equation~(\ref{main3})  must be replaced  by
\begin{gather}
\nabla^2 \Psi = s \bra  \hat{\delta\dot\phi} \ket = (s/a) \bra \hat{\pi} \ket.
\label{SemiEE}
\end{gather}
At this point, one  can  clearly observe that
 if  the  state of the  quantum f\/ield  is in the vacuum state,  the metric
  perturbations  vanish and thus  the space-time is homogeneous and isotropic.

The  proposal  is  based on  the consideration of  a self induced collapse, which
  we  take  to  operate in close analogy with  a ``measurement'' (but  evidently, with no external  measuring apparatus  or observer   involved). This  leads us to want  to work
 with  Hermitian operators, which in ordinary quantum mechanics are the ones susceptible of direct measurement.
 Therefore,  we must  separate  both $\hat y_{\nk} (\eta)$ and $\hat \pi_{\nk}
(\eta)$ into their real and imaginary parts $\hat y_{\nk} (\eta)=\hat y_{\nk}{}^R
(\eta) +i \hat y_{\nk}{}^I (\eta)$ and $\hat \pi_{\nk} (\eta) =\hat \pi_{\nk}{}^R
(\eta) +i \hat \pi_{\nk}{}^I (\eta)$, where
\begin{gather*}
\hat y_{\nk}{}^{R,I} (\eta) = \frac{1}{\sqrt{2}} \big( y_k (\eta) \hat{a}_{\nk}^{R,I} + y_k^* (\eta) \hat{a}_{\nk}^{{R,I} \dag}\big),  \qquad \hat \pi_{\nk}{}^{R,I} (\eta) = \frac{1}{\sqrt{2}} \big( g_k (\eta) \hat{a}_{\nk}^{R,I} + g_k^* (\eta) \hat{a}_{\nk}^{R,I \dag}\big),
\end{gather*}
with
$
\hat{a}_{\nk}^R \equiv \frac{1}{\sqrt{2}} (\hat{a}_{\nk} + \hat{a}_{-\nk} )$, $\hat{a}_{\nk}^I \equiv \frac{-i}{\sqrt{2}} ( \hat{a}_{\nk} - \hat{a}_{-\nk}).
$
Then the operators $\hat y_{\nk}^{R, I} (\eta)$ and $\hat
\pi_{\nk}^{R, I} (\eta)$ are hermitian\footnote{Note however that he commutation relations of the real and imaginary creation and annihilation operators are however non-standard:
$[\hat{a}_{\nk}^R, \hat{a}_{\nk'}^{R,\dag}] = \hbar L^3 (\delta_{\nk,\nk'} + \delta_{\nk,-\nk'})$, $[\hat{a}_{\nk}^I, \hat{a}_{\nk}^{I,\dag} ] = \hbar L^3 (\delta_{\nk,\nk'} - \delta_{\nk,-\nk'})$,
with all other commutators vanishing. This is known to indicate that the operators corresponding to $\nk$ and $-\nk$ are identical in the real case and identical up to a sign in the imaginary case.}.

So  far,  the  treatment   is  similar  to the standard  one,   except in that only the scalar  f\/ield is  being   treated at the quantum level, while   the  metric  f\/luctuation is  treated  classically.  At this point, it is   worthwhile  to   emphasize  that the  vacuum state   def\/ined  by $ \ann_{\nk}{}
^{R,I} |0\ket =0$  is  100\%  translational  and rotationally invariant,  as  can be  seen from the fact that it is
annihilated  by the   generators of translations   and rotations   (i.e.\ by the linear and the angular momentum operators, $\hat{\vec P}$ and $\hat{\vec L}$, respectively).

Next we  need to  specify  in more detail  the   modeling of  the  collapse.  Then,   take into account that  after   the collapse has  taken place,  we must  consider the  continuing  evolution  of the  expectation values of the   f\/ield  variables
 until  the end of inf\/lation, and eventually  up to the  last scattering hypersurface (in fact,  if we  want to actually   compare  our analysis with  the detailed  observations,  we must  evolve  also through  the  reheating  period  and  through the decoupling era up  to  today's  Universe. This,  however,  is   normally  taken into account  through the use of appropriate transfer functions, and  we  will assume that  the  same  procedure can be implemented  after the present analysis).

We will further assume that the collapse is somehow analogous to an imprecise measurement\footnote{An imprecise measurement of an observable, is one in which one does not end with an exact eigenstate  of that observable but  rather with a state which is  only peaked around the eigenvalue. Thus, we could consider measuring a  certain particle's position and momentum, ending  up with a state characterized by wave packet with both position and momentum def\/ined to a limited extent, and which,  of course, does not  entail a conf\/lict with Heisenberg's uncertainty bound.}
of the
operators $\hat y_{\nk}
^{R, I}
(\eta)$ and $\hat \pi_{\nk}
^{R, I}
(\eta)$.
Now, we will specify the rules according to which the collapse happens.
Again, at this point our criteria will be simplicity and naturalness.
What we have to describe is the state $|\Theta\rangle$ after the
collapse.
 It turns out that, for the goals at hand, all we need to specify  is the  expectation  values  immediately  after the collapse
as this  determines   the  expectation value of the f\/ield and momentum  operator for the mode  $\nk$ at all times after the collapse.
That  is using the result in  the evolution equations  for the  expectation values (i.e.\ using  Ehrenfest's theorem), one  obtains
$\bra {\hat{y}_{\nk}{}^{R,I} (\eta)} \ket$ and
 $\bra {\hat{\pi}_{\nk}{}^{R,I} (\eta)} \ket $ for the   state  that resulted from the collapse, for all later times.

 As we  indicated,  at this  point we  assume that  after the  collapse, the expectation values of the f\/ield and momentum operators  in each mode
  will  be related to the uncertainties  of the  pre-collapse  state (recall that the  expectation  values  in the vacuum state  are  zero). In the vacuum state, $\hat{y}_{\nk}$ and
$\hat{\pi}_{\nk}$ individually are distributed according to Gaussian
wave functions centered at~0 with spread $\fluc{\hat{y}_{\nk}}_0$ and
$\fluc{\hat{\pi}_{\nk}}_0$, respectively.

  We  might   consider  various possibilities   for the detailed  form  of this collapse. Thus, for   their generic form,
  associated with the  ideas  above, we write:
\begin{gather}
\bra {\hat{y}_{\nk}^{R,I}
(\eta^c_k)} \ket_\Theta= \lambda_y x
^{R,I}
_{\nk,y}
\sqrt{\fluc{\hat{y}^{R,I}_{\nk}}_0}
=\lambda_y x^{R,I}_{\nk,y}|y_k(\eta^c_k)|\sqrt{\hbar L^3/2},
\label{momentito}
\\
\bra {\hat{\pi}_{\nk}{}^{R,I}
(\eta^c_k)}\ket_\Theta
= \lambda_\pi x
^{R,I}
_{\nk,\pi}\sqrt{\fluc{\pyRI_{\nk}}
_0} ,
=\lambda_\pi x^{R,I}_{\nk,\pi}|g_k(\eta^c_k)|\sqrt{\hbar L^3/2},
\label{momentito1}
\end{gather}
where $\tc$ represents the \emph{time of collapse} for each mode. The random variables $x_{\nk,y}^{R,I}$,
$x_{\nk,\pi}^{R,I}$ are  selected randomly from a distribution which, in principle, might be non-Gaussian,  i.e.\ dif\/ferent modes could present particular correlations (we will elaborate on this in the next section).  At this point, we must emphasize that our Universe corresponds
to a single realization of these random variables, and thus  each of these quantities
$ x^{R,I}_{\nk,y}$, $ x^{R,I}_{\nk,\pi}$ have a  single specif\/ic value.
The  two parameters    $\lambda_y$ and $\lambda_\pi$  are  real numbers we use  to characterize the dif\/ferent collapse schemes, e.g.: i)~$\lambda_y=0$, $\lambda_\pi=1$, ii)~$\lambda_y=\lambda_\pi=1$ (which we call the ``symmetric scheme''). It is clear that one can devise  many other models  of  collapse,   a good  fraction of them  can be  described   within the  scheme  above,   while others require  a  slightly  modif\/ied treatment~\cite{Adolfo}.  Still,
there are surely   many other  possibilities which  we  have not  even  thought about  and which might require  drastically  modif\/ied  formalisms.

 The explicit expressions for the $\bra \hat{y}_{\nk}^{R,I} (\eta) \ket_\Theta$ and $\bra \hat{\pi}_{\nk}^{R,I} (\eta) \ket_\Theta$ are then
\begin{gather}
\bra \hat{y}_{\nk}^{R,I} (\eta) \ket_\Theta  =  \left[ \frac{\cos D_k}{k} \left( \frac{1}{k\eta} - \frac{1}{z_k} \right) + \frac{\sin D_k}{k} \left(\frac{1}{k\eta z_k} + 1 \right) \right]  \bra \hat{\pi}_{\nk}^{R,I} (\tc) \ket_\Theta \nonumber \\
\hphantom{\bra \hat{y}_{\nk}^{R,I} (\eta) \ket_\Theta  =}{}
 +  \left( \cos D_k - \frac{\sin D_k}{k\eta} \right)  \bra \hat{y}_{\nk}^{R,I} (\tc) \ket_\Theta, \label{expecyeta}
\\
\label{expecpieta}
\bra \hat{\pi}_{\nk}^{R,I} (\eta) \ket_\Theta  = \left( \cos D_k +\frac{\sin D_k}{z_k} \right) \bra \hat{\pi}_{\nk}^{R,I} (\tc) \ket_\Theta  - k \sin D_k  \bra \hat{y}_{\nk}^{R,I} (\tc) \ket_\Theta,
\end{gather}
where $D_k (\eta) \equiv k\eta - z_k$ and $z_k \equiv k\tc$.

   With  this information at  hand  we can now  compute the perturbations of the metric  after the collapse of   all the modes\footnote{In fact,  we need only  be concerned with the relevant modes, those  that  af\/fect the observational quantities in a relevant  way. Modes that  have  wavelengths that  are   either too large or too  small are irrelevant in this  sense.}.

\subsection{The observable quantities}

Now, we must put together our semi-classical  description  of the gravitational
degrees of freedom and the quantum mechanics description of the inf\/laton f\/ield.  We recall that this entails the semi-classical version of
the perturbed Einstein's equation that, in our case, leads to equation~(\ref{SemiEE}).
The Fourier components  at  the conformal time $ \eta$ are given by:
\begin{gather}
 \Psi_{\nk} ( \eta) = -\sqrt{\frac{\epsilon}{2}} \frac{H_I \hbar}{M_P k^2} \bra {\hat{\pi}_{\nk}{}(\eta)}  \ket,
\label{SemiEEK}
\end{gather}
where we used that during inf\/lation $ s \equiv 4 \pi G \dot{\phi_0} = \sqrt{\epsilon/2}a H_I / M_P$, with $M_P$ the reduced Planck's mass $M_P^2 \equiv \hbar^2/(8 \pi G)$.  The expectation value depends  on  the  state of the quantum f\/ield, therefore,   as  we have  already noted, prior to the collapse, we have  $\Psi_{\nk} ( \eta)  =0$, and the space-time is  still homogeneous and isotropic at the corresponding scale. However, after the collapse  takes place,  the state of the f\/ield  is a dif\/ferent state
with new expectation values which  generically  will not vanish, indicating that after this time the Universe becomes anisotropic and in-homogeneous at the corresponding scale.
 We now can reconstruct the space-time value of the Newtonian potential using
$\Psi(\eta,\vec{x})=
 \frac{1}{L^{3}}\sum_{\vec k}\ e^{i\vec{k}\cdot\vec{x}} \Psi_{\nk}
(\eta)
$,
to extract the quantities of observational interest.

 In order to make contact with the observations, we shall relate the expression \eqref{SemiEEK} for the evolution of
the Newtonian potential during the early phase of accelerated expansion, to the small anisotropies observed in the
 temperature of the cosmic microwave background radiation, $\delta T(\theta,\varphi)/T_0$ with $T_0\approx 2.725 K$ the temperature average.
They are considered as the f\/ingerprints of the small perturbations pervading the Universe at the time of decoupling, and undoubtedly any model for the origin of the seeds of cosmic structure should account for them.

The observational data is   described  in terms of the coef\/f\/icients  $\alpha_{lm}$
of the multipolar series expansion
\begin{gather}\label{expansion.alpha}
\frac{\delta T}{T_0}(\theta,\varphi)=\sum_{lm}\alpha_{lm}Y_{lm}(\theta,\varphi),\qquad
\alpha_{lm}=\int \frac{\delta T}{T_0}(\theta,\varphi)Y^*_{lm}(\theta,\varphi)d\Omega .
\end{gather}
Here $\theta$ and $\varphi$ are the coordinates on the celestial two-sphere, with $Y_{lm}(\theta,\varphi)$ the spherical harmonics.
The dif\/ferent multipole numbers $l$ correspond to dif\/ferent angular scales; low $l$ to large scales and high $l$ to small scales. At large angular scales ($l \lesssim 20$) the anisotropies in the CMB arise due to the Sachs--Wolfe ef\/fect.
That ef\/fect relates the anisotropies in the temperature observed today on the celestial sphere to the inhomogeneities in the Newtonian potential on the last scattering surface:
$\frac{\delta T}{T_0} (\theta,\varphi) = \frac{1}{3} \Psi (\eta_D, \vec{x}_D)$,
where $\eta_D$ is the conformal
time of decoupling which lies in the matter-dominated epoch, and $\vec{x}_D = R_D (\sin \theta \sin \varphi, \sin \theta \cos \varphi, \cos \theta)$, with $R_D$ the radius of the
last scattering surface. Furthermore, using $\Psi(\eta,\vec{x})=
 \frac{1}{L^{3}}\sum_{\vec k}\ e^{i\vec{k}\cdot\vec{x}} \Psi_{\nk}
(\eta)$ and  $e^{i \vec{k} \cdot \vec{x}_D} = 4 \pi \sum_{lm} i^l j_l (kR_D) Y_{lm} (\theta, \varphi) Y_{lm}^* (\hat{k})$, the expression (\ref{expansion.alpha}) for $\alpha_{lm}$ can be rewritten in the form
\begin{gather}\label{alm2}
\alpha_{lm} = \frac{4 \pi i^l}{3L^3}   \sum_{\nk} j_l (kR_D) Y_{lm}^* (\hat{k}) \Delta (k)
 \Psi_{\vec{k}} (\eta_R),
\end{gather}
with $j_l (kR_D)$ the spherical Bessel function of order $l$.
 The transfer function $\Delta(k)$ represents the evolution of the Newtonian potential from the end of inf\/lation $\eta_R$  to the last scattering surface at the time of decoupling $\eta_D$, i.e.\  $\Psi_{\vec{k}}(\eta_D)=\Delta (k) \Psi_{\vec{k}} (\eta_R)$.
 We  will  be  ignoring this  aspect   from this point onward,   despite the  fact that this  transfer functions   are  behind  the famous acoustic peaks, the most noteworthy  feature of the CMB  spectrum,   as  they  relate to   aspects of plasma physics that are well understood and thus  uninteresting from our point of view. This  will    mean that,  from our point of view,  the observational  spectrum  would  have  such  features removed  before comparing with our results (this is   in the same  spirit  that one removes the  uninteresting imprint of our galaxy  on the  observations   of the CMB, or the  dipole associated with our peculiar motion). This would be  equivalent  to assume that the  observed  spectrum f\/its   well with  the   f\/lat Harrison--Zel'dovich  spectrum and  setting $\Delta (k)  =1$, and thus  setting  for our purposes here  $\Psi_{\vec{k}}(\eta_D)= \Psi_{\vec{k}} (\eta_R)$.

Substituting \eqref{expecpieta} in \eqref{SemiEEK} and using \eqref{momentito}, \eqref{momentito1} gives
\begin{gather}
\Psi_{\nk} (\eta_R)  =  \frac{- (L\hbar)^{3/2} \sqrt{ \epsilon} H_I }{2 \sqrt{2} M_P  k^{3/2}} \left[ \lambda_\pi \left( \cos D_k  + \frac{D_k }{z_k} \right) \big(x_{\nk,\pi}^R + i x_{\nk,\pi}^I\big) \right.\nonumber \\
\left. \phantom{\Psi_{\nk} (\eta_R)  =}{}   -  \lambda_y \sin D_k  \left( 1 + \frac{1}{z_k^2} \right)^{1/2}  \big(x_{\nk,y}^R + i x_{\nk,y}^I\big) \right].\label{psirandom}
\end{gather}
where $z_k \equiv k \tc$, we note that the quantity $D_k$ is being evaluated at $\eta_R$, i.e.\ $D_k (\eta_R) = k\eta_R - z_k$.

Finally using, \eqref{psirandom} in \eqref{alm2} yields:
\begin{gather}
\alpha_{lm}  = - \frac{ \pi i^l \hbar^{3/2} \sqrt{ 2 \epsilon} H_I }{3L^{3/2} M_P} \sum_{\nk} \frac{j_l (k R_D)}{k^{3/2}} Y^*_{lm} (\hat{k}) \left[ \lambda_\pi \left( \cos D_k + \frac{\sin D_k  }{z_k} \right) \big(x_{\nk,\pi}^R + i x_{\nk,\pi}^I\big) \right.\nonumber \\
\left. \phantom{\alpha_{lm}  =}{} - \lambda_y \sin D_k \left(1+ \frac{1}{z_k^2} \right)^{1/2} \big(x_{\nk,y}^R + i x_{\nk,y}^I\big) \right].\label{almcol}
\end{gather}
We  will  further   simplify  the  treatment by noting that the value of  $-k\eta_R $  is  exceedingly small for the modes of interest. Thus,  this  will be ignored in comparison to $z_k$, i.e.\   we  will be assuming that  we  can
make the approximation $D_k \to -z_k$.

It is worthwhile to mention that the relation of $\alpha_{lm}$ with the Newtonian potential, as obtained
 in \eqref{almcol} within the collapse framework has  no analogue in the  usual  treatments of the subject.
 It  provides us  with a  clear  identif\/ication   of the  aspects  of the  analysis  where  the ``randomness''  is
  located.  In this case,  it resides  in the  randomly   selected values $ x_{\nk,y}^{\textrm{R,I}}$, $ x_{\nk,\pi}^{\textrm{R,I}}$
  that appear in the   expressions of the collapses associated  with  each of the modes. Here, we also f\/ind a clarif\/ication of
   how is it that, in spite  of  the intrinsic  randomness,  we  can  make  any prediction at all. The  individual  complex quantities
   $\alpha_{lm} $  correspond to  large sums of complex  contributions,  each  one  having a  certain randomness,  but leading
    in  combination to a  characteristic  value,  in  just  the same  way as a random walk that is made  up   of multiple steps. In other words,
     the justif\/ication for the use of statistics in our approach is: The quantity~$\alpha_{lm}$ is the sum of contributions from the
      collection of modes, each contribution being a random number leading to what is in ef\/fect a
      sort of ``two-dimensional random walk'', which total displacement corresponds to the observational quantity.
      Nothing  like  this can be found in the most popular accounts, in which the issues  we have been focussing on  are
       hidden  in a maze  of often  unspecif\/ied  assumptions and unjustif\/ied  identif\/ications~\cite{Short}.

Finally, let us discuss the implications of our proposal regarding the tensorial perturbations.
 The analysis we present  is for the sake of completeness, as we have already presented it  in previous works~\cite{multiples, Us}, and is not the main subject of this article.

\looseness=-1
The  view we take, regarding the relation between the metric (classical) description of gravity and the quantum
 matter degrees of freedom, has  a direct implication on the estimates for the production of gravitational waves (from inf\/lation).
 The idea is that the mechanism that generates the actual  the f\/luctuations is tied to  on the quantum uncertainties for the scalar f\/ield. Due to some unknown
  quantum gravitational ef\/fect, the state of each mode of the scalar f\/ield collapses. After the collapse, the density inhomogeneities and anisotropies  feed into the gravitational
degrees of freedom resulting in perturbations (scalar, vector and tensor) of the metric. Nevertheless, the metric itself does not induce a quantum gravitational
collapse. Therefore, as the scalar f\/ield does not act as a source for the gravitational
tensor modes~-- at least not at the order considered here~-- the tensor modes cannot be
excited. Thus, the collapse scheme naturally predicts\footnote{However, it is worthwhile pointing out that such a
conclusion is directly tied to our underlying approach that favors the semi-classical Einstein's equations augmented with a collapse proposal as a way to deal with the gravity quantum
interface faced in the current problem. It is of course conceivable, although seems harder to understand in a wider
context (see the discussion in Section~8 of~\cite{Sudarsky2011}), that a collapse might be incorporated into a setting where both the
gravitation and scalar f\/ield perturbations are simultaneously treated at the quantum level. If the latter happened to
be the correct approach, something that would be possible to ascertain when we have a fully satisfactory theory of
quantum gravity, our conclusion about the tensor modes would be modif\/ied. } a zero~-- or at least a strongly suppressed~-- amplitude of gravitational waves into the CMB.

\section{Modif\/ied statistical features in the primordial spectrum}\label{NG}

In this section, we will show how the statistical aspects of the collapse model might induce non-trivial features in the primordial spectrum.
This is a novel aspect and contrasts with what has been   done  in the previous  approaches, where the dif\/ferent modes that contribute to the primordial spectrum are  considered to be uncorrelated.

Let us start by noting that according to (\ref{almcol}) all the modes contribute to $\alpha_{lm}$
with a  complex number. If we had  the outcomes   characterizing  each of the individual collapses  we  could, of course,  predict  the  value of these quantities. However, we
have  at this point no other  access to such information  than the observational quantities  $
   \alpha_{lm}$ themselves.

     We  hope  to be able to say something about these, but  doing so   requires the consideration  of further  hypothesis  regarding the statistical  aspects  of  the  physics
      behind the collapse as well as the conditions previous to them.

 As  is  generally the case  with  random walks, one can not hope to  estimate the direction of the  f\/inal displacement.  However,  one might  say something about its estimated
 magnitude.  It is for that reason  we  will be  focussing on estimating the  most likely   value of  the  magnitude $|\alpha_{lm} |^2$. We can make  some progress,  for instance,  by
  making some assumption  allowing us  to   regard the  specif\/ic  outcomes characterizing our Universe as  a typical member  of some  hypothetical  ensemble of Universes.

For example, we are interested in  estimating  the  most likely  value of  the magnitude
of $| \alpha_{lm}|^2$ above,  and    in  such  hypothetical  ensemble  we might hope that  it comes very close to our single  sample.
    It is  worthwhile emphasizing  that,  for each $l$ and $m$,  we have one single   complex number characterizing the actual observations (and thus the  real  Universe  we inhabit).
    For a given $l$ for instance, we should  avoid  confusing  ensemble averages    with averages of   such  quantities  over the $2l+1$ values of $m$.
    The  other universes   in the ensemble  are  just constructs  of our imagination and there is nothing in our theories that  would  indicate that they are real.

We can  simplify things  even further  by  taking   the ensemble average $  \overline{|\alpha_{lm}|^2}$ (recall that the bar represents an  ensemble
   average)   and  identifying it with the most likely value  of the quantity, and   needless is to say that  these  two notions are dif\/ferent in many types of ensembles.
    However, let us, for the moment, ignore this issue  and   assume the   identity of  those   two  values.  We   thus have
\begin{gather}
\label{alm8}
 | \alpha_{lm}|^2_{\textrm{ML}}  = \overline{| \alpha_{lm}|^2}  =  \frac{2 \pi^2 \hbar^3 \epsilon H_I}{9L^3 \mpl^2  }
\sum_{\vec{k}, \vec{k'}}  \frac{j_l (kR_D)}{k^{3/2}} \frac{j_l (k'R_D)}{k^{'3/2}} Y_{lm}^* (\hat{k}) Y_{lm} (\hat{k'})   \\
{}
\times   \Big[\lambda_\pi M(z_k) M(z_{k'}) ( \overline{x_{\nk,\pi}^R x_{\nk',\pi}^R} + \overline{ x_{\nk,\pi}^I x_{\nk',\pi}^I}) + \lambda_y^2 N(z_k) N(z_{k'}) ( \overline{x_{\nk,y}^R x_{\nk',y}^R} + \overline{x_{\nk,y}^I x_{\nk',y}^I} )\Big], \nonumber
\end{gather}
with
\[
M(z_k) \equiv \cos z_k - \frac{\sin z_k  }{z_k},   \qquad N(z_k) \equiv \sin z_k  \bigg(1+ \frac{1}{z_k^2} \bigg)^{1/2},
\]
and we have  used the assumption that the four groups of
random variables $x_{\nk,y}^R$, $x_{\nk,y}^I$, $x_{\nk,\pi}^R$, $x_{\nk,\pi}^I$ are completely uncorrelated (leading thus to the cancelation of the cross terms in~\eqref{alm8}).

As mentioned in Section \ref{ssc}, our  ref\/ined  approach  suggests that, within each set of random variables, there would  be a correlation between the modes
$\nk$ and $2\nk$ (and also of course between~$\nk$ and~$\nk/2$).
This correlation would induce a ``small'' departure  from the  normal  probability distribution function
 of the random variables, where the ``smallness'' is characterized by the introduction of a parameter $\varepsilon$ (do not confuse~$\varepsilon$ with the slow-roll parameter~$\epsilon$).
 That is, we would assume that the ensemble average of the product of two random variables is given by
\begin{gather}\label{equisr}
\overline{x_{\nk,i}^R x_{\nk',i}^R} = \delta_{\nk,\nk'} + \delta_{\nk,-\nk'} + \varepsilon \big( \delta_{\nk,2 \nk'} + \delta_{\nk,-2 \nk'} + \delta_{2 \nk,\nk'} +\delta_{-2\nk,\nk'} \big),
\\
\label{equisi}
\overline{x_{\nk,i}^I x_{\nk',i}^I} = \delta_{\nk,\nk'} - \delta_{\nk,-\nk'} + \varepsilon \big( \delta_{\nk,2 \nk'} - \delta_{\nk,-2 \nk'} + \delta_{2 \nk,\nk'} -\delta_{-2\nk,\nk'} \big),
\end{gather}
with $i = y,\pi$.
Note that  this  suggests  also  many more complex  and higher order correlations  between various sets  of variables,  but we  will  neglect those  higher order ef\/fects  here, in part  because
we do not have  a specif\/ic  self-content mold  for those at this point.
We recall that the random variables corresponding to~$\nk$ and~$-\nk$ are not independent. Using~\eqref{equisr} and~\eqref{equisi} in~\eqref{alm8} we obtain
\begin{gather}
 | \alpha_{lm}|^2_{\textrm{ML}}   =  \frac{4 \pi^2 \hbar^3 \epsilon H_I}{9L^3 \mpl^2  } \sum_{\nk,\nk'} \frac{j_l(k R_D)}{k^{3/2}} \frac{j_l(k'R_D)}{k^{'3/2}} Y_{lm}^* (\hat{k}) Y_{lm} (\hat{k}') \nonumber \\
 \phantom{| \alpha_{lm}|^2_{\textrm{ML}}   =}{}
   \times  \left[ \lambda_\pi^2 M(z_k) M(z_{k'})  + \lambda_y^2 N(z_k) N(z_{k'}) \right] [ \delta_{\nk,\nk'} + \varepsilon (\delta_{\nk,2\nk'} + \delta_{2\nk,\nk'} )].\label{alm9}
\end{gather}

As  we  have found in previous  works,  the  data  indicates that,  to  a good  degree of accura\-cy~\cite{Adolfo},
the assumption that $z_k$ is independent of $k$, i.e.\ the time of collapse goes as $\tc \propto k^{-1}$, is a good approximation. Therefore, assuming $z_k = z= {\rm const}$,  \eqref{alm9} takes the form
\begin{gather*}
 | \alpha_{lm}|^2_{\textrm{ML}}   =   \frac{4 \pi^2 \hbar^3 \epsilon
H_I C(z) }{9L^3 M_{P}^{2}  } \Bigg\{ \sum_{\nk} \frac{|Y_{lm} (\hat{k}
) |^2}{k^3} \Big[ j_l^2 (kR_D) + \varepsilon \big(  2^{3/2} j_l (kR_D)
j_l (k R_D/2 ) \\
\phantom{| \alpha_{lm}|^2_{\textrm{ML}}   =}{} +   2^{-3/2} j_l (kR_D) j_l (2 k R_D )  \big) \Big] \Bigg\},
\end{gather*}
where $C(z) \equiv \lambda_\pi^2 M^2(z) + \lambda_y^2 N^2(z)$. Now we write the sum as an integral by noting that the allowed values of the
components of $\nk$ are separated by $\Delta k_i = 2\pi /L$, that is
\begin{gather*}%\label{alm10}
 | \alpha_{lm}|^2_{\textrm{ML}}   =  \frac{ \hbar^3 \epsilon H_I C(z) }{18 \pi \mpl^2  } \left[ \int_0^\infty \frac{dk}{k}  j_l^2 (kR_D) \right.\nonumber \\
\left.\hphantom{| \alpha_{lm}|^2_{\textrm{ML}}   =}{}
 +  \varepsilon  \int_0^\infty \frac{dk}{k} \left( 2^{3/2} j_l (k R_D) j_l (k R_D / 2) + 2^{-3/2} j_l (k R_D) j_l (2k R_D)   \right) \right],
\end{gather*}
where in the previous expression we performed the integral over the angular part of~$\nk$.

Rewriting the remaining integrals, we f\/ind
\begin{gather*}
 | \alpha_{lm}|^2_{\textrm{ML}} = \frac{\hbar^3 \epsilon H_I  C(z)}{18 \pi \mpl^2  } \left[ \int_0^\infty \frac{dx}{x} j_l^2(x)
 + \frac{9}{2^{3/2}} \varepsilon \int_0^\infty \frac{dx}{x} j_l(x) j_l(2x) \right],
\end{gather*}
where we use a change of variable $x\equiv kR_D$. Evaluating the integrals of the products of spherical Bessel's functions gives the f\/inal result
\begin{gather}\label{alm12}
 | \alpha_{lm}|^2_{\textrm{ML}}  = \frac{ \hbar^3 \epsilon H_I C(z)}{18 \pi \mpl^2  } \bigg[ \frac{1}{2l(l+1)}
+ \varepsilon  \frac{9 \sqrt{\pi} \Gamma(l)}{2^{7/2+l} \Gamma(l+3/2)} \textrm{ }_2F_1 (l,-1/2  ; l +3/2, 1/4 ) \bigg],
\end{gather}
where $\Gamma(l)$ is the Gamma function, and $_2F_1 (\alpha,\beta; \gamma,z)$ is the hypergeometric function\footnote{The hypergeometric function is def\/ined as $_2F_1 (\alpha,\beta; \gamma,z) \equiv \frac{\Gamma(\gamma)}{\Gamma(\beta) \Gamma(\gamma- \beta)} \int_0^1 \frac{t^{\beta-1} (1-t)^{\gamma - \beta-1}}{(1-tz)^\alpha} dt$.}.

Now, as
 \eqref{alm12} does not depend on $m$ it is clear that the expectation of the quantity of observational
 interest, i.e.\ the angular spectrum  $l(l+1)C_l \equiv l(l+1)(2l+1)^{-1} \sum_m |\alpha_{lm}|^2$, is just $|\alpha_{lm}|^2$ and thus this quantity must be compared  with:
\begin{gather}\label{cl1}
l(l+1)C_l = {\cal A} [ 1 + \varepsilon G(l)],
\end{gather}
where we have def\/ined
\begin{gather*}%\label{gl}
G(l) \equiv \frac{l(l+1) 9 \sqrt{\pi} \Gamma(l)}{2^{5/2 + l} \Gamma(l+3/2)} \; {}_2F_1 (l,-1/2  ; l +3/2, 1/4 ),
\end{gather*}
and $\mathcal{A}$   is a normalization constant that  is of no interest to us here. Clearly,  if $\varepsilon = 0$,
 which corresponds to the case that the probability distribution functions  for the variables $x_{\nk,i}^R$, $x_{\nk',i}^R$ and
  $x_{\nk,i}^I$, $x_{\nk',i}^I$ are  uncorrelated,  random, and Gaussian (see \eqref{equisr} and \eqref{equisi}), then we recover the
  standard f\/lat spectrum, i.e.\ $l(l+1)C_l$ is independent of~$l$.  The departure of the f\/lat spectrum has  a  very specif\/ic  signature that,
   in principle, can be searched for  observationally. There are of course  many  other possibilities  of unexpected  correlations and  each one  will  have a particular signature.
    \begin{figure}[t]
    \centering
    \includegraphics{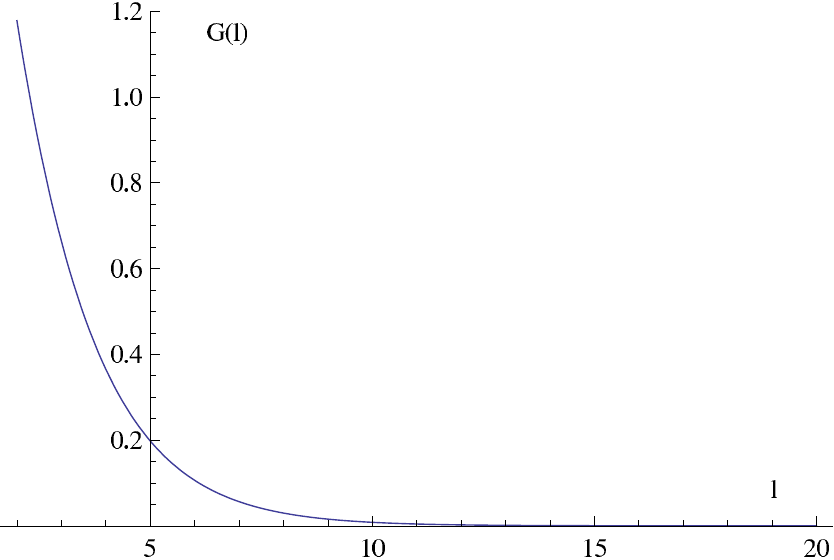}
    \caption{$G(l)$ for $l \in [2,20]$.  The deviation from the f\/lat spectrum will only af\/fect the low multipoles.}\label{graf1}
    \end{figure}

The behavior of $G(l)$ for  $2 \leq l \leq 20$ is shown in Fig.~\ref{graf1}. As we can see, the ef\/fect is stronger for low $l$ and it decreases
in a nearly exponential fashion for large $l$ (we have   made  a  direct  conf\/irmation of   this  fact).
The next step, of a more rigorous analysis, would be to consider
the acoustic oscillations, due to plasma physics, which play an important role after the end of the inf\/lationary regime. Then, we
could use the recent observational data available (e.g.~WMAP7), and thus put some constraints on the parameter $\varepsilon$.

Recall, however,  that the result~\eqref{cl1}  we obtained depends on the  behavior  of another parameter we have introduced, namely the
time of collapse~$\tc$. Let us  recall, that in order to recover  the detailed form  of the  spectrum, we  have had to make   the  additional
  assumption that $\tc \propto 1/k$ (i.e.~$z_k$ is independent of~$k$). At this point we  only have a very   heuristic  proposal
    that seems to account for this assumption (see \cite{Us}, in terms  of an  argument that attempts to connect that pattern with certain
      ideas  of  Penrose's and Diosi's linking the collapse of the wave function with  aspects of quantum gravity). In
        fact, the assumption that $\tc \propto 1/k$ seems  necessary to recover something  close to  the observed  spectral shape,
        as has  already   been noted in previous works~\cite{Adolfo, Us}. The point is that, in order to constrain
        the values of $\varepsilon$ and $z_k$, we should perform an analysis such as the one presented in~\cite{claudia}.
        That work,   consisted of a detailed  statistical analysis testing some specif\/ic  ``collapse schemes'',  and establishing
         constraints on the free parameters of the  modes  considered. The analysis was carried out comparing the  details predictions
          obtained  by  a modif\/ied  CMB fast code incorporating out models, and  recent data from the CMB, including the 7-yr data release
          of the WMAP~\cite{WMAP7} and the matter power spectrum measured employing the Luminous Red Galaxies by the Sloan Digital Sky Survey~\cite{SDSS}.
           A~similar analysis for the results of the present paper lies outside the scope of this work. However, our main result of this work, i.e.\ the pattern noted in~\eqref{cl1},
            indicates that the appearance of the correlations between the  modes $\nk$ and $2\nk$;  af\/fecting  the statistics of the
             random variables $x_{\nk,i}^R$, $x_{\nk',i}^R$ and $x_{\nk,i}^I$, $x_{\nk',i}^I$, and  thus our estimates for  the observational quantities.
              It seems  that looking for  such specif\/ic    pattern of  deviation in the  form of CMB  spectrum, would  be  a  straight forward (but  highly  resource  consuming) computational endeavor.

Finally, let us end this section by noting a heuristic relation between  our result \eqref{cl1} and   the so called ``parametric resonance''  ef\/fect in quantum optics.
 The set up
involves a parametric nonlinear crystal excited  by an intense laser beam at
angular frequency $\omega_p = 2\omega$.  The interesting fact is  that a weak signal beam at angular frequency
$\omega$ is spontaneously  generated.  In a  particular  setting, the nonlinear crystal  can   mix a   second  input
signal $\omega_s=\omega$  with that  resulting from the  pump, and this produces an idler beam at angular
frequency $\omega$.   The nonlinear process either amplif\/ies
or de-amplif\/ies the  excitation  of   the signal  modes depending on the  phases. Thus, the
output f\/ield is  the   result of  delicate  quantum interference resulting from a highly correlated  state.
In short,  the nonlinear  nature of the  crystal's  response leads to photons of one frequency being converted into photons
with a dif\/ferent one, leading to the  appearance in the spectrum of the beam
of important connections   between  the components  with a given frequency and its  f\/irst harmonics.
The connection with our result \eqref{cl1} would lead  us to view  the  gravitational response to the
 collapse of matter f\/ields, as described by the SSC scheme, playing the role of  the ``non-linear crystal''.
 That is, the collapse seems to  correlate the mode $\nk$ the modes $n \nk$, ($n$~an integer),  in the same sense as the crystal
 correlates the angular frequency of the signal and the pump. The term proportional to $\varepsilon$ in \eqref{cl1}
 would correspond to the modif\/ication on the amplitude in the previous example.

\section{Discussion}\label{discusion}

The  work presented  here is  a continuation of a  long standing program  aimed  at addressing and  proposing a possible
 cure for a set of shortcomings  that af\/f\/lict the  standard accounts for the emergence of the seeds of  cosmic  structure during inf\/lation.
  The program has  revealed that, although  the motivating problems  can  be characterized as conceptual dif\/f\/iculties, it is
   intimately  connected    with what is  usually called the measurement problem in quantum mechanics. This problem,  which as
   we  have argued,  becomes  essentially insurmountable within the standard  quantum theory  in the cosmological context (see \cite{Short}).
   Nevertheless, one  is  led   surprisingly to  expect observational  consequences  associated  with some details of the physical
   unknown mechanisms that lie  beyond quantum theory and
   our current  understanding.  That is,  by  parameterizing  some  aspects of  this  enigmatic process, we  might
   actually  use  observations  as a  guides  in learning something about  them. In previous works,  we have seen, for
    instance,  that the details of the ``collapse  schemes''  can  modif\/ied  the primordial  f\/luctuation spectral shape in   very  particular ways \cite{Us}.
     We  have  argued that  we  can use  the  observations to place  limits   on the times of  collapse  of the various modes \cite{Adolfo}.

We  have discussed that   the semiclassical setting should be the  appropriate framework to  deal with inf\/lationary perturbations. This is  simply
  because the full  quantum theory is likely to  be one  where  the   space-time notions  are not  present at all (due in part  by the  problem
  of time  that af\/f\/licts  canonical  approaches to quantum gravity such as LQG). As has  been   argued  elsewhere   in that  setting,   there  would not be
   tensor modes  exited  at  f\/irst order in perturbation theory and   thus  no  primordial gravity waves of  a  magnitude that could  be  observed in the forceable future.

Moreover, in the approach we have  been using in this and previous  works, it becomes quite transparent on how  the statistical aspects are related to each
one of  the observational quantities. That is,  the observable $\alpha_{lm}$ is related to the random variables $x_{\nk}$'s
through a complex ``random walk'', while  the value of $|\alpha_{lm}|^2$ corresponds to the
squared ``length'' of the random walk. This random walk is associated to a particular realization of a physical quantum process (i.e.\  associated
with one of the  multiple  possibilities  characterized by the collapse of the inf\/laton's wave-function). However, as  we  have  access to just
 one such realization~-- the set of  random walks corresponding to our own Universe~-- the most natural
 course of  action (but certainly not the only one) is to take the  ensemble average value of the length of the possible random walks,
 which corresponds to $\overline{|\alpha_{lm}|^2}$, to be identif\/ied with the most likely value, i.e.\ with $|\alpha_{lm}|_{\textrm{ML}}^2$; and this in
 turn, is to  be  compared  with the $|\alpha_{lm}|^2$ of our observable Universe. Of course,  our Universe  is to be  identif\/ied  with  one  realization
 of the random variables, and as  such, the resulting picture has  an interpretational  transparency that seems  hard  to f\/ind in the standard accounts.

The  clarif\/ication mentioned above  has  allowed  us to take a  fresh approach in addressing the possibility of novel statistical features and the
 manner in  which those  could be searched for, as  discussed  in~\cite{Susana}. In the present work we have discussed  a  novel
   possibility  for  nonstandard   statistics  uncovered  by our approach. This novel aspect  is tied to the   possibility  that  non trivial
    statistical features  would arise as  a result of the  of  the collapse  process itself. Namely that   the collapse  of one  of the  modes  (say  the mode $\vec k$)
    might inf\/luence  the  tendency  in  which  the state  of a related
    mode  (say the mode $ 2 \vec k$) to collapse in  a certain  manner.  We can heuristically  view   this possibility
    as  ref\/lecting a  modif\/ication of the statistical aspects of the  state of  one mode  as  the result of the collapse of another.

We  noted  that the deviation from the standard prediction  would be particularly relevant   for  the f\/irst multipoles: It is thus  interesting to speculate
whether  this ef\/fect could be linked to the issues raised in~\cite{Starkman1,Starkman2}. In those works, it has been argued that there are specif\/ic large-scale anomalies
 in the temperature map of anisotropies of the CMB for the low  modes.
According to~\cite{Starkman2}, the observations disagree markedly with the predictions of the standard theory. By that  as it might,
it seems,  that our result \eqref{cl1}, which was motivated by purely conceptual issues,  shares some  of the essential features of those  observations, i.e.\ a departure of the
standard~-- f\/lat~-- spectrum for the lowest $l$. Evidently, at this stage of the analysis we cannot present more
quantitative results, however, we believe that the previous discussion serves to show the potential behind our proposal.

In any  event, we have seen that  such kind of  modif\/ications  have   particular signatures  which  could,  in principle,  be   searched  for  in the   data,  leading as  we  have said,  either  to  interesting  discoveries,  or  at least to the setting of bounds on the type of ef\/fect we have  been considering.

\subsection*{Acknowledgements}

The work of GL and DS was supported  in part by the    CONACYT grant
No 101712. GL acknowledges f\/inancial support by CONACYT postdoctoral grant. DS was  also supported  by the PAPIIT-UNAM  grant IN107412. We thank the referees for useful suggestions.

\pdfbookmark[1]{References}{ref}
\LastPageEnding


\begin{thebibliography}{99}
\footnotesize\itemsep=0pt

\bibitem{Guth_3}
Albrecht A., Steinhardt P.J., Cosmology for grand unif\/ied theories with
  radiatively induced symmetry breaking, \href{http://dx.doi.org/10.1103/PhysRevLett.48.1220}{\textit{Phys. Rev. Lett.}} \textbf{48}
  (1982), 1220--1223.

\bibitem{Bartolo}
Bartolo N., Komatsu E., Matarrese S., Riotto A., Non-{G}aussianity from
  inf\/lation: theory and observations, \href{http://dx.doi.org/10.1016/j.physrep.2004.08.022}{\textit{Phys. Rep.}} \textbf{402} (2004),
  103--266, \href{http://arxiv.org/abs/astro-ph/0406398}{astro-ph/0406398}.

\bibitem{Bassi}
Bassi A., Ghirardi G.C., Dynamical reduction models, \href{http://dx.doi.org/10.1016/S0370-1573(03)00103-0}{\textit{Phys. Rep.}}
  \textbf{379} (2003), 257--426, \href{http://arxiv.org/abs/quant-ph/0302164}{quant-ph/0302164}.

\bibitem{Starkman1}
Copi C.J., Huterer D., Schwarz D.J., Starkman G.D., Large angle anomalies in
  the CMB, \href{http://dx.doi.org/10.1155/2010/847541}{\textit{Adv. Astron.}} \textbf{2010} (2010), 847541, 17~pages,
  \href{http://arxiv.org/abs/1004.5602}{arXiv:1004.5602}.

\bibitem{Adolfo}
De~Un\'anue A., Sudarsky D., Phenomenological analysis of quantum collapse as
  source of the seeds of cosmic structure, \href{http://dx.doi.org/10.1103/PhysRevD.78.043510}{\textit{Phys. Rev.~D}} \textbf{78}
  (2008), 043510, 15~pages, \href{http://arxiv.org/abs/0801.4702}{arXiv:0801.4702}.

\bibitem{colapsosmuk}
Diez-Tejedor A., Le\'on G., Sudarsky D., The collapse of the wave function in
  the joint metric-matter quantization for inf\/lation, \href{http://arxiv.org/abs/1106.1176}{arXiv:1106.1176}.

\bibitem{Alberto}
Diez-Tejedor A., Sudarsky D., Towards a formal description of the collapse
  approach to the inf\/lationary origin of the seeds of cosmic structure,
  \href{http://arxiv.org/abs/1108.4928}{arXiv:1108.4928}.

\bibitem{Diosi}
Di\'osi L., A universal master equation for the gravitational violation of
  quantum mechanics, \href{http://dx.doi.org/10.1016/0375-9601(87)90681-5}{\textit{Phys. Lett.~A}} \textbf{120} (1987), 377--381.

\bibitem{Diosi1}
Di\'osi L., Models for universal reduction of macroscopic quantum f\/luctuations,
  \href{http://dx.doi.org/10.1103/PhysRevA.40.1165}{\textit{Phys. Rev.~A}} \textbf{40} (1989), 1165--1174.

\bibitem{Gambini1}
Gambini R., Porto R.A., Pullin J., Fundamental decoherence from relational time
  in discrete quantum gravity: {G}alilean covariance, \href{http://dx.doi.org/10.1103/PhysRevD.70.124001}{\textit{Phys. Rev.~D}}
  \textbf{70} (2004), 124001, 8~pages, \href{http://arxiv.org/abs/gr-qc/0408050}{gr-qc/0408050}.

\bibitem{Gambini}
Gambini R., Porto R.A., Pullin J., Realistic clocks, universal decoherence, and
  the black hole information paradox, \href{http://dx.doi.org/10.1103/PhysRevLett.93.240401}{\textit{Phys. Rev. Lett.}} \textbf{93}
  (2004), 240401, 3~pages, \href{http://arxiv.org/abs/hep-th/0406260}{hep-th/0406260}.

\bibitem{GRW}
Ghirardi G.C., Rimini A., Weber T., Unif\/ied dynamics for microscopic and
  macroscopic systems, \href{http://dx.doi.org/10.1103/PhysRevD.34.470}{\textit{Phys. Rev.~D}} \textbf{34} (1986), 470--491.

\bibitem{Guth_1}
Guth A.H., Inf\/lationary universe: a~possible solution to the horizon and
  f\/latness problems, \href{http://dx.doi.org/10.1103/PhysRevD.23.347}{\textit{Phys. Rev.~D}} \textbf{23} (1981), 347--356.

\bibitem{Mukhanov1990_3}
Guth A.H., Pi S.Y., Fluctuations in the new inf\/lationary universe,
  \href{http://dx.doi.org/10.1103/PhysRevLett.49.1110}{\textit{Phys. Rev. Lett.}} \textbf{49} (1982), 1110--1113.

\bibitem{Guth_4}
Guth A.H., Pi S.Y., Quantum mechanics of the scalar f\/ield in the new
  inf\/lationary universe, \href{http://dx.doi.org/10.1103/PhysRevD.32.1899}{\textit{Phys. Rev.~D}} \textbf{32} (1985),
  1899--1920.

\bibitem{Mukhanov1990_4}
Halliwell J.J., Hawking S.W., Origin of structure in the universe,
  \href{http://dx.doi.org/10.1103/PhysRevD.31.1777}{\textit{Phys. Rev.~D}} \textbf{31} (1985), 1777--1791.

\bibitem{Mukhanov1990_1}
Hawking S.W., The development of irregularities in a single bubble inf\/lationary
  universe, \href{http://dx.doi.org/10.1016/0370-2693(82)90373-2}{\textit{Phys. Lett.~B}} \textbf{115} (1982), 295--297.

\bibitem{Guth_5}
Hawking S.W., Moss I.G., Fluctuations in the inf\/lationary universe,
  \href{http://dx.doi.org/10.1016/0550-3213(83)90319-X}{\textit{Nuclear Phys.~B}} \textbf{224} (1983), 180--192.

\bibitem{isham}
Isham C.J., Canonical quantum gravity and the problem of time, in Integrable
  Systems, Quantum Groups, and Quantum Field Theories ({S}alamanca, 1992),
  \textit{NATO Adv. Sci. Inst. Ser. C Math. Phys. Sci.}, Vol.~409, Kluwer Acad.
  Publ., Dordrecht, 1993, 157--287, \href{http://arxiv.org/abs/gr-qc/9210011}{gr-qc/9210011}.

\bibitem{Komatsu}
Komatsu E., Hunting for primordial non-{G}aussianity in the cosmic microwave
  background, \href{http://dx.doi.org/10.1088/0264-9381/27/12/124010}{\textit{Classical Quantum Gravity}} \textbf{27} (2010), 124010,
  26~pages, \href{http://arxiv.org/abs/1003.6097}{arXiv:1003.6097}.

\bibitem{Komatsu2001}
Komatsu E., The pursuit of non-gaussian f\/luctuations in the cosmic microwave
  background, Ph.D. thesis, Tohoku University, 2001, \href{http://arxiv.org/abs/astro-ph/0206039}{astro-ph/0206039}.


\bibitem{claudia}
Landau S.J., Scoccola C.G., Sudarsky D., Cosmological constraints on
  non-standard inf\/lationary quantum collapse models, \href{http://arxiv.org/abs/1112.1830}{arXiv:1112.1830}.

\bibitem{WMAP7}
Larson D., Dunkley J., Hinshaw G. et~al., Seven-year Wilkinson microwave
  anisotropy probe (WMAP) observations: power spectra and WMAP-derived
  parameters, \href{http://dx.doi.org/10.1088/0067-0049/192/2/16}{\textit{Astrophys.~J. Suppl.}} \textbf{192} (2011), 16, 19~pages,
  \href{http://arxiv.org/abs/1001.4635}{arXiv:1001.4635}.

\bibitem{multiples}
Le{\'o}n G., De~Un{\'a}nue A., Sudarsky D., Multiple quantum collapse of the
  inf\/laton f\/ield and its implications on the birth of cosmic structure,
  \href{http://dx.doi.org/10.1088/0264-9381/28/15/155010}{\textit{Classical Quantum Gravity}} \textbf{28} (2011), 155010, 22~pages,
  \href{http://arxiv.org/abs/1012.2419}{arXiv:1012.2419}.

\bibitem{Susana}
Le\'{o}n G., Landau S.J., Sudarsky D., Quantum origin of the primordial f\/luctuation
  spectrum and its statistics, \href{http://arxiv.org/abs/1107.3054}{arXiv:1107.3054}.

\bibitem{Gabriel}
Le{\'o}n G., Sudarsky D., The slow-roll condition and the amplitude of the
  primordial spectrum of cosmic f\/luctuations: contrasts and similarities of the
  standard account and the `collapse scheme', \href{http://dx.doi.org/10.1088/0264-9381/27/22/225017}{\textit{Classical Quantum
  Gravity}} \textbf{27} (2010), 225017, 23~pages, \href{http://arxiv.org/abs/1003.5950}{arXiv:1003.5950}.

\bibitem{Liguori}
Liguori M., Sefusatti E., Fergusson J.R., Shellard E.P.S., Primordial
  non-Gaussianity and bispectrum measurements in the cosmic microwave
  background and large-scale structure, \href{http://dx.doi.org/10.1155/2010/980523}{\textit{Adv. Astron.}} \textbf{2010}
  (2010), 980523, 64~pages, \href{http://arxiv.org/abs/1001.4707}{arXiv:1001.4707}.

\bibitem{Guth_2}
Linde A.D., Coleman--{W}einberg theory and the new inf\/lationary universe
  scenario, \href{http://dx.doi.org/10.1016/0370-2693(82)90086-7}{\textit{Phys. Lett.~B}} \textbf{114} (1982), 431--435.

\bibitem{Lyth}
Lyth D.H., Liddle A.R., The primordial density perturbation: cosmology,
  inf\/lation and the origin of structure, Cambridge University Press, Cambridge,
  2009.

\bibitem{Mukhanov}
Mukhanov V., Physical foundations of cosmology, \href{http://dx.doi.org/10.1017/CBO9780511790553}{Cambridge University Press}, New
  York, 2005.

\bibitem{Mukhanov1990}
Mukhanov V.F., Chibisov G.V., Quantum f\/luctuation and nonsingular universe,
  \textit{JETP Lett.} \textbf{33} (1981), 532--535.

\bibitem{Mukhanov1990_5}
Mukhanov V.F., Feldman H.A., Brandenberger R.H., Theory of cosmological
  perturbations, \href{http://dx.doi.org/10.1016/0370-1573(92)90044-Z}{\textit{Phys. Rep.}} \textbf{215} (1992), 203--333.

\bibitem{Padmanabhan}
Padmanabhan T., Structure formation in the universe, Cambridge University
  Press, New York, 1993.

\bibitem{Pearle}
Pearle P.M., Dynamical wave function collapse: could it have cosmological
  consequences?, \href{http://arxiv.org/abs/0710.0567}{arXiv:0710.0567}.


\bibitem{Penrose1}
Penrose R., On gravity's role in quantum state reduction, \href{http://dx.doi.org/10.1007/BF02105068}{\textit{Gen.
  Relativity Gravitation}} \textbf{28} (1996), 581--600.

\bibitem{Penrose}
Penrose R., The emperor's new mind. Concerning computers, minds, and the laws
  of physics, The Clarendon Press, Oxford University Press, New York, 1989.


\bibitem{Us}
Perez A., Sahlmann H., Sudarsky D., On the quantum origin of the seeds of
  cosmic structure, \href{http://dx.doi.org/10.1088/0264-9381/23/7/008}{\textit{Classical Quantum Gravity}} \textbf{23} (2006),
  2317--2354, \href{http://arxiv.org/abs/gr-qc/0508100}{gr-qc/0508100}.

\bibitem{SDSS}
Reid B.A., Percival W.J., Eisenstein D.J. et~al., Cosmological constraints
  from the clustering of the sloan digital sky survey DR7 luminous red
  galaxies, \href{http://dx.doi.org/10.1111/j.1365-2966.2010.16276.x}{\textit{Mon. Not. Roy. Astron. Soc.}} \textbf{404} (2010), 60--85,
  \href{http://arxiv.org/abs/0907.1659}{arXiv:0907.1659}.

\bibitem{Starkman2}
Starkman G.D., Copi C.J., Huterer D., Schwarz D., The oddly quiet universe: how
  the CMB challenges cosmology's standard model, \href{http://arxiv.org/abs/1201.2459}{arXiv:1201.2459}.

\bibitem{Guth}
Starobinsky A.A., A~new type of isotropic cosmological models without
  singularity, \href{http://dx.doi.org/10.1016/0370-2693(80)90670-X}{\textit{Phys. Lett.~B}} \textbf{91} (1980), 99--102.

\bibitem{Mukhanov1990_2}
Starobinsky A.A., Dynamics of phase transition in the new inf\/lationary universe
  scenario and generation of perturbations, \href{http://dx.doi.org/10.1016/0370-2693(82)90541-X}{\textit{Phys. Lett.~B}} \textbf{117}
  (1982), 175--178.

\bibitem{Sudarsky:2007tp}
Sudarsky D., A signature of quantum gravity at the source of the seeds of
  cosmic structure?, \href{http://dx.doi.org/10.1088/1742-6596/67/1/012054}{\textit{J.~Phys. Conf. Ser.}} \textbf{67} (2007), 012054,
  6~pages, \href{http://arxiv.org/abs/gr-qc/0701071}{gr-qc/0701071}.

\bibitem{Sudarsky2011}
Sudarsky D., Can we learn something about the quantum/gravity interface from
  the primordial f\/luctuation spectrum?, \href{http://dx.doi.org/10.1142/S0218271811019165}{\textit{Internat.~J. Modern Phys.~D}}
  \textbf{20} (2011), 821--838.

\bibitem{Short}
Sudarsky D., Shortcomings in the understanding of why cosmological
  perturbations look classical, \href{http://dx.doi.org/10.1142/S0218271811018937}{\textit{Internat.~J. Modern Phys.~D}}
  \textbf{20} (2011), 509--552, \href{http://arxiv.org/abs/0906.0315}{arXiv:0906.0315}.

\bibitem{Sudarsky:2006zx}
Sudarsky D., The seeds of cosmic structure as a door to new physics,
  \href{http://dx.doi.org/10.1088/1742-6596/68/1/012029}{\textit{J.~Phys. Conf. Ser.}} \textbf{68} (2007), 012029, 12~pages,
  \href{http://arxiv.org/abs/gr-qc/0612005}{gr-qc/0612005}.

\bibitem{Wald1994}
Wald R.M., Quantum f\/ield theory in curved spacetime and black hole
  thermodynamics, \textit{Chicago Lectures in Physics}, University of Chicago Press,
  Chicago, IL, 1994.

\bibitem{Weinberg}
Weinberg S., Cosmology, Oxford University Press, Oxford, 2008.

\bibitem{Yadav}
Yadav A.P.S., Wandelt B.D., Primordial non-Gaussianity in the cosmic microwave
  background, \href{http://dx.doi.org/10.1155/2010/565248}{\textit{Adv. Astron.}} \textbf{2010} (2010), 565248, 27~pages,
  \href{http://arxiv.org/abs/1006.0275}{arXiv:1006.0275}.

\end{thebibliography}
\end{document}